# Enabling News Consumers to View and Understand Biased News Coverage: A Study on the Perception and Visualization of Media Bias

Timo Spinde[1,3], Felix Hamborg[1], Karsten Donnay[1,2], Angelica Becerra[1], Bela Gipp[1,3]
University of Konstanz, Germany {first.last}@uni-konstanz.de
University of Zurich, Switzerland, {first.last}@uzh.ch
University of Wuppertal, Germany {last}@uni-wuppertal.de

## ABSTRACT

Traditional media outlets are known to report political news in a biased way, potentially affecting the political beliefs of the audience and even altering their voting behaviors. Many researchers focus on automatically detecting and identifying media bias in the news, but only very few studies exist that systematically analyze how theses biases can be best visualized and communicated. We create three manually annotated datasets and test varying visualization strategies. The results show no strong effects of becoming aware of the bias of the treatment groups compared to the control group, although a visualization of hand-annotated bias communicated bias instances more effectively than a framing visualization. Showing participants an overview page, which opposes different viewpoints on the same topic, does not yield differences in respondents' bias perception. Using a multilevel model, we find that perceived journalist bias is significantly related to perceived political extremeness and impartiality of the article.

## 1 Introduction

News articles serve as a highly relevant source of information on current topics and salient political issues [1]. News coverage is not just the communication of facts; on the contrary, news articles put facts into context and transport specific opinions. Hence, how the news covers a topic or issue can decisively impact public debates and affect our collective decision making [2]. Slanted tonality, word choice, and other forms of media bias may have a large impact on individuals' perceptions of societal issues. The severity of biased news coverage is amplified further by the fact that regular news consumers are typically not fully aware of its degree and scope [3].

While previous research projects aim to manually or automatically identify media bias [4,5], so far, not much research has been conducted on how to visualize instances of bias most effectively. The literature describes ways to present news, e.g., news aggregators [4,6], but currently lacks measuring effectiveness and efficiency of how to visualize and communicate single instances of media bias to enable news consumers to become aware of bias and aid them in understanding its effects on their perception of news topics.

We present the results of a prototypical user study, in which we test the effectiveness of communicating bias-related news characteristics using different visualization types and components. Our experiments include tests on an overview level, e.g., showing multiple news topics, related articles, and an aggregated measure of slanted language, as well as on an article level, e.g., showing a selected article's text and in-text bias instances. After an extensive literature review on user-related variables that may affect users' perceptions of bias, such as their political background, we devise and test a questionaire for assessing the perception of bias and four visualizations: two control group visualizations (bias agnostic) as well as two treatment visualizations (bias aware). The contributions of this paper are:

C1. Analysis of the influence of participant related factors as to understanding biased coverage.
C2. Analysis of the efficacy of mitigating echo chamber effects by presenting users different perspectives on the same topic.
C3. Comparison of the efficacy of bias agnostic and bias aware visualization types as to enabling news consumers to understand the presence and degree of biased coverage.

## 2 Related Work

We first introduce user-related factors that may influence how users perceive bias. Then, we give an overview of existing bias visualizations.

Which factors influence if and how readers perceive media bias in the news? Druckman et al. showed the importance of asking study participants about their political background before showing them articles [7]. Eveland et al. related bias to various variables, such as age, education, income, gender, political orientation, and political involvement [1]. They showed that perceived hostile bias is strongly dependent on the participants' political stance, political involvement, and political discussions with like-minded individuals. Alessio et al. showed how readers were more likely to designate material opposing their own viewpoints as biased. They also summarized categories to measure articles, e.g., biased / balanced, accurate / mistaken, complete / lacking / detailed or fair / one-sided [8]. The results of a study on the effects of editorial bias by Ardevol et al. indicates that perceived media bias is associated with the amount of individual daily news usage [9].

Lastly, Kause et al. showed that participants with lower levels of numeracy skills generally perceived problems as more severe and were also more likely to be concerned generally [3].

While only few studies test the effectiveness of bias identification systems and compare bias-aware with bias-agnostics visualizations, to our knowledge, no study to date has investigated individual visualization components as to their effectiveness in communicating biases. User studies by Hamborg et al. [4] and Park et al. [6,10] confirm that bias-aware visualizations in general help users to become aware of bias, compared to baseline visualizations that were bias-agnostic. An et al. gave a prototypical visualization of media bias on Twitter [11]. While they showed how their model could help people receive balanced news information, they also deemphasized "the potential benefit of such political diversity because not everyone prefers to receive diverse political opinions" [11]. A user study on NewsCube 2, a crowdsourcing system for framing in the news, showed that exposing opposing viewpoints on one topic can lead "readers to read more articles covering different aspects" [6] and help them to develop more balanced views.

## 3 METHODOLOGY

Our study employs a conjoint analysis [12] to test how visualizations can improve users' understanding or awareness of media bias in news articles. In our design, we show a fully randomized selection of visualization variants to each participant and then ask a series of questions regarding the perception of bias in the articles viewed. We base our selection of variables for this prototypical study, as shown in Table 1, on the literature review results summarized in Section 2. Future research will include more variables and visualization types (see Section 5).

**Table 1: Participant-related variables presumed to influence bias perception.**

| Variable | Category | Source |
| --- | --- | --- |
| Gender, age, education, religion, residence | Demogr. | [1] |
| Political orientation, esp. party affiliation | Pol. bg. | [13] |
| Opinion on pre-selected, political topics | Pol. bg. | [14] |
| Engaging in political discussions | Pol. bg. | [1] |
| Belief in hostile media | Pol. bg. | [1] |

A full set of questions and all visualizations mentioned in the following can be inspected online (see Section 5).

The workflow of the study is as follows. First, participants answer the previously described questions on their background. Second, we randomly assign respondents (1) to an order of three pre-selected topics (to mitigate any influence of the order in which topics are viewed [15], (2) to seeing the overview visualization or not (to measure contribution C2), and (3) to one of the four article visualizations (to measure C3). The design of the overview follows allsides.com, per topic showing three articles that are representative of three political categories, as depicted in Figure 1.

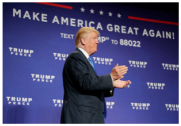

**Noncitizen voter fraud allegations**

According to a 2012 Pew Research Center (Pew) report, which Donald Trump has cited as proof of voter fraud, millions of voter registration records were out of date as people were either deceased or had moved.

| From the left | From the center | From the right |
|---|---|---|
| **FOLLOWING TRUMP VOTER FRAUD ALLEGATIONS, CLAIM THAT 5.7 MILLION NONCITIZENS VOTED IS WRONG** | **Report: Trump commission did not find widespread voter fraud** | **Hillary Says Voter Registration Cost Her the Election** |
| President Donald Trump's unfounded allegations that millions voted illegally in 2016 is back in the news, with... | PORTLAND, Maine (AP) — The now-disbanded voting integrity commission launched by the Trump ... | President Trump was criticized last year when he tweeted out an accusation that he would have won the popular vote were it ... |
| **Source: Tampa Bay Times** | **Source: Associated Press** | **Source: National Review** |
| **Strength of slanted language bar:** | **Strength of slanted language bar:** | **Strength of slanted language bar:** |

**Figure 1: Excerpt from the news overview page.**

Then, we show the texts of three articles (one article after another) to the participants, whereby we visualize the articles differently: the control group sees the text (1) *plain*, i.e., as it would be shown in a news aggregator or on a news website, or (2) with visually highlighted phrases that represent the facts most important to an *article's event*. The treatment groups see the text enhanced with either (3) visually highlighted *framing effects*, or (4) visually highlighted annotations of *biased or unbalanced language*. An excerpt from this fourth variant is shown in Figure 2. In variant 4, we also show reasons why the text is biased. Variant 4 aims to represent instances of bias found using the currently most effective form of bias analysis, i.e., we conducted an inductive content analysis with 6 coders. Variant 3 aims to represent what the state-of-the-art in automated bias identification is able to detect (cf. [15–17]), e.g., target-dependent sentiment classification, a basic yet effective way to catch the effects of biased coverage. Since we want to test the visualization effectiveness but not the underlying detection technique, the variant 3 instances also stem from a manual annotation with six coders.

We gathered the annotations used within the visualizations for each of the articles by a brief questionnaire, where three students per article and type of annotation, i.e., framing (3) and biased language (4), marked text phrases based on their judgment. Another three students checked their results manually, before we integrated all results into one common set of annotations for each article, by discarding annotations that were not found by at least two of the three students in both groups. To facilitate the appearance of bias, we selected a publicly controversial and politically polarizing topic (immigrant voter Fraud Allegations and immigration restrictions) as well as one topic related to the fake university in Farmington.

**Figure 2: Excerpt from one article with visually highlighted annotations of biased language.**

Fifth, after viewing each article, we asked participants a series of questions, mainly seeking to measure if and how strongly they became aware of the presence or absence of bias within the article. We asked (1) two control questions, e.g., about the article's content, to verify that they had read the article. For example, for the immigration topic, we asked: "How many illegal immigrants are believed to might have voted, according to the article?" To further understand whether participants rather agreed or disagreed with the opinion voiced in the article, we asked (2) how much participants agreed with a polarizing statement from the current article and how much they believed that the public agrees with that statement. For example, for the first article, we asked how much the participant agreed with the statement: "Trump has made repeated claims about massive voter fraud and election rigging." Most importantly, we asked (3) how the participants estimated the degree of the bias of the article's author, how politically extreme they perceived the news article, and how impartial and one-sided they thought it was in dealing with the actual issue. The questions were indirectly asking for bias perception, as we assumed a strong emotional and personal effect when asking for bias directly. The existing literature did not show a common perspective on how to ask for bias. We will address this again in future work. After answering the questions, our survey respectively moved to the next article, exposing each respondent to three articles overall. In both groups, i.e., treatment and control group, the type of visualization and order of articles were randomized.

## 4 RESULTS

Participants of the study were US Turkers on Amazon Mechanical Turk (MTurk). To reduce the number of low-quality answers, we accepted only Turkers with MTurk's Masters' qualification [18]. Overall, 123 workers completed the study. We performed a manual quality analysis afterward and discarded the low-quality results of one Turker.

We find that for all of our variables (C1), random effects show a high variation between the three articles. While our set of experimental variables did not lead to significant differences in means, a multilevel model shows that perceived journalist bias was directly and significantly related to perceived political extremeness and impartiality of the article. The model can also be inspected online (see Section 5). We did not find any influence of the time at which participants saw an article on bias perception, e.g. as first or last one of the three we showed them.

Exposure to divergent perspectives in the overview visualization, e.g., article excerpts reprefsenting the political spectrum of the same topic, does not significantly alter the awareness of media bias in the articles viewed afterward (C2). While this is, on the one hand, unintuitive (people will become aware of other perspectives), on the other hand, it is in line with findings by An et al., i.e., readers are resistant against views different from their view [11]. Forcing users to view different perspectives may increase hesitance even further. We think that future research should focus on raising interest in users to view opposing perspectives (see Section 5).

The results show no strong effects of becoming aware of the bias of the treatment groups compared to the control group (C3), which is partially in line with prior work on echo chambers (see also C2) and that news readers tend to prefer reading articles matching with their views [11]. However, we notice some effects of different visualizations. For example, Figure 3 shows that the hand-annotated bias visualization (first column), which reveals biased vs. neutral language, most effectively communicates bias instances to users. Based on a significance level of 10%, our multilevel model from C1 confirms this. The framing visualization (second) yields slight improvements and the important-fact visualization (third) no improvements compared to the control group. Lastly, we find that readers can determine if and how much an article is biased, e.g., impartial or politically extreme, since different users usually agree regarding their rating on the same article. Figure 3 also shows strong differences between the articles shown in the study, e.g., article A1 (red) was deemed more biased as to multiple variables, such as political extremeness and journalist's unfairness, than A2 (green).

## 5 CONCLUSION AND FUTURE WORK

We present the results of a user study on the effectiveness of communicating slanted bias coverage and, more specifically, individual instances of media bias in news articles to news consumers using different visualizations. Specifically, we investigate three parts that may influence the perception of slanted coverage: readers' background (contribution C1), viewing a bias aware news overview (C2), and different visualizations for reading an article (C3). While on the one hand, the study finds no statistically significant factors influencing bias perception in users, on the other hand, we find several indicative factors that we plan to investigate in the future in more detail.

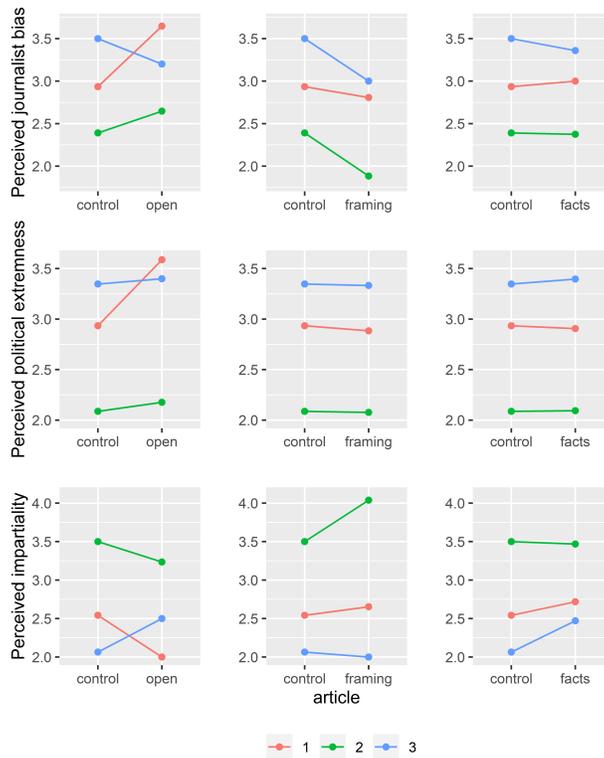

**Figure 3: Perceived level of political extremeness, fair perspective and impartiality (each in one row) on a scale from 1 (least) to 5 (most), comparing the visualizations with the control group (columns).**

For C1, we find that random effects show strong variation dependent on the article. Perceived political extremeness, journalist bias, and impartiality were closely and significantly related. We confirmed readers' aversion against other views [11] by our overview test (C2). We plan to investigate alternatives to forcing users to viewing different perspectives, e.g., indicating different word choices of one fact within an article. We will also investigate whether there exist differences in bias perception when forcedly seeing a specific article after an overview page or when there exists a free article choice, which could be a major difference to the news aggregator by Park et al. [10]. We find that participants become aware of bias when presented with bias-aware visualizations (C3): visualizing annotations stemming from a manual content analysis are most, followed by highlighting targets as to their sentiment. In the future, we plan to devise individual visualization components that are either aware or agnostic of bias [16]. We also plan to check differences between asking for bias directly and indirectly. Lastly, we plan to automate the underlying annotation process [15,17,19] and expand the approach to other languages [20].

We publish the survey materials, including questionnaires, articles, visualizations, data and results freely at: https://zenodo.org/record/3627995#.Xi3uOyNCeUk

## ACKNOWLEDGMENTS

This work was supported by a grant from the Heidelberg Academy of Sciences and Humanities. It has also been funded by the Hanns-Seidel-Foundation supported by the Federal Ministry of Education and Research of Germany.

## REFERENCES

[1] Jr. Eveland, P. William, and D. V. Shah. "The impact of individual and interpersonal factors on perceived news media bias". In: *Political Psychology 24*. 2003, pp. 101–117. DOI: 10.1111/0162-895X.00318
[2] D. Eil and J.M. Rao. "The good news-bad news effect: asymmetric processing of objective information about yourself ". In: *American Economic Journal: Microeconomics 3*. 2011, pp. 114–138. DOI: 10.1257/mic.3.2.114
[3] A. Kause, T. Townsend and W. Gaissmaier. "Framing Climate Uncertainty: Frame Choices Reveal and Influence Climate Change Beliefs". In: *Weather, Climate, and Society 11*. 2019, pp. 199–215. DOI: 10.1175/WCAS-D-18-0002.1
[4] F. Hamborg, N. Meuschke, and B. Gipp. "Matrix-based news aggregation: exploring different news perspectives". In: *Proceedings of the 17th ACM/IEEE Joint Conference on Digital Libraries*. 2017, pp. 69–78. DOI: 0.1007/s00799-018-0239-9
[5] T. Ogawa, Q. Ma, and M. Yoshikawa. "News bias analysis based on stakeholder mining". In: *IEICE transactions on information and systems 94*. 2011, pp. 578–586. DOI: 10.1587/transinf.E94.D.578
[6] S. Park, S. Kang, S. Chung, and J. Song." NewsCube: delivering multiple aspects of news to mitigate media bias". In: *Proceedings of the SIGCHI Conference on Human Factors in Computing Systems*. 2009, pp. 443-452. DOI: 10.1145/1518701.1518772

[7] J.N. Druckman and M. Parkin. "The impact of media bias: How editorial slant affects voters" In: *The Journal of Politics 67*. 2005, pp. 1030–1049. DOI: 10.1111/j.1468-2508.2005.00349.x

[8] D. D'Alessio. "An experimental examination of readers' perceptions of media bias". In: *Journalism & mass communication quarterly 80*. 2003, pp. 282–294. DOI: 10.1177/107769900308000204

[9] A. Ardèvol-Abreu and H. Gil De Zúñiga. „Effects of editorial media bias perception and media trust on the use of traditional, citizen, and social media news". In: *Journalism & mass communication quarterly 94*. 2017, pp. 703–724. DOI: 10.1177/1077699016654684

[10] S. Park, J.K. M. Ko, H. Choi, and J. Song. "NewsCube 2.0: an exploratory design of a social news website for media bias mitigation". In: *Workshop on Social Recommender Systems*. 2011.

[11] J. An, M. Cha, K. Gummadi, J. Crowcroft, and D. Quercia. "Visualizing media bias through Twitter". In: *Sixth International AAAI Conference on Weblogs and Social Media*. 2012.

[12] J. Hainmueller, D. Hopkins, and T. Yamamoto. "Causal inference in conjoint analysis: Understanding multidimensional choices via stated preference experiments". In: *Political Analysis 22*. 2014, pp. 1–30. DOI: 10.1093/pan/mpt024

[13] J. Graham, J. Haidt, and B. Nosek. "Liberals and conservatives rely on different sets of moral foundations". In: *Journal of personality and social psychology 96*. 2009, pp. 1029-1046. DOI: doi.org/10.1037/a0015141

[14] E. Lee. "That's not the way it is: How user-generated comments on the news affect perceived media bias". In: *Journal of Computer-Mediated Communication 18*. 2012, pp. 32–45. DOI: 10.1111/j.1083-6101.2012.01597.x

[15] F. Hamborg, K. Donnay, and B.Gipp. "Automated identification of media bias in news articles: an interdisciplinary literature review". In: *International Journal on Digital Libraries 20*. 2019, pp. 391–415. DOI: 10.1007/s00799-018-0261-y

[16] F. Hamborg, A. Zhukova, K. Donnay, and B. Gipp. "Newsalyze: Enabling News Consumers to Understand Media Bias". In: *Proceedings of the ACM/IEEE Joint Conference on Digital Libraries (JCDL)*. 2020, pp. 179–187. DOI: 10.1145/3383583.3398561

[17] F. Hamborg, A. Zhukova, and B. Gipp. "Illegal Aliens or Undocumented Immigrants? Towards the Automated Identification of Bias by Word Choice and Labeling". In: *International Conference on Information*. 2019, pp. 179–187. DOI: 10.1007/978-3-030-15742-5_17

[18] M. Lovett, S. Bajaba, M. Lovett, and M. Simmering. "Data quality from crowdsourced surveys: A mixed method inquiry into perceptions of amazon's mechanical turk masters". In: *Applied Psychology 67*. 2018, pp. 339–366. DOI: 10.1111/apps.12124

[19] F. Hamborg, A. Zhukova, and B. Gipp. "Automated Identification of Media Bias by Word Choice and Labeling in News Articles". In: *ACM/IEEE Joint Conference on Digital Libraries (JCDL)*. 2019, pp. 196–205. DOI: 10.1109/jcdl.2019.00036

[20] T. Spinde, F. Hamborg, and B. Gipp. "An Integrated Approach to Detect Media Bias in German News Articles". In: *Proceedings of the ACM/IEEE Joint Conference on Digital Libraries (JCDL)*. 2020. DOI: 10.1145/3383583.3398585